\documentstyle[11pt]{article}
\catcode`\@=11
\long\def\@makefntext#1{
\protect\noindent \hbox to 3.2pt {\hskip-.9pt
$^{{\ninerm\@thefnmark}}$\hfil}#1\hfill}                
\def\@makefnmark{\hbox to 0pt{$^{\@thefnmark}$\hss}}  
\def\ps@myheadings{\let\@mkboth\@gobbletwo
\def\@oddhead{\hbox{}
\rightmark\hfil\ninerm\thepage}
\def\@oddfoot{}\def\@evenhead{\ninerm\thepage\hfil
\leftmark\hbox{}}\def\@evenfoot{}
\def\sectionmark##1{}\def\subsectionmark##1{}}
\renewcommand{\thefootnote}{\fnsymbol{footnote}}
\newcounter{sectionc}\newcounter{subsectionc}\newcounter{subsubsectionc}
\renewcommand{\section}[1] {\vspace*{0.6cm}\addtocounter{sectionc}{1}
\setcounter{subsectionc}{0}\setcounter{subsubsectionc}{0}\noindent
        {\normalsize\bf\thesectionc. #1}\par\vspace*{0.4cm}} 
\renewcommand{\subsection}[1] {\vspace*{0.6cm}\addtocounter{subsectionc}{1}
        \setcounter{subsubsectionc}{0}\noindent
    {\normalsize\it\thesectionc.\thesubsectionc. #1}\par\vspace*{0.4cm}}
\renewcommand{\subsubsection}[1]
{\vspace*{0.6cm}\addtocounter{subsubsectionc}{1}
\noindent {\normalsize\rm\thesectionc.\thesubsectionc.\thesubsubsectionc.
    #1}\par\vspace*{0.4cm}}

\newcounter{appendixc}
\newcounter{subappendixc}[appendixc]
\newcounter{subsubappendixc}[subappendixc]

\renewcommand{\appendix}[1] {\vspace*{0.6cm}
        \refstepcounter{appendixc}
        \setcounter{figure}{0}
        \setcounter{table}{0}
        \setcounter{equation}{0}
        \renewcommand{\thefigure}{\Alph{appendixc}.\arabic{figure}}
        \renewcommand{\thetable}{\Alph{appendixc}.\arabic{table}}
        \renewcommand{\theappendixc}{\Alph{appendixc}}
        \renewcommand{\theequation}{\Alph{appendixc}.\arabic{equation}}
        \noindent{\bf Appendix \theappendixc #1}\par\vspace*{0.4cm}}

\renewenvironment{thebibliography}[1]
    {\begin{list}{\arabic{enumi}.}
    {\usecounter{enumi}\setlength{\parsep}{0pt}
\setlength{\leftmargin 1.25cm}{\rightmargin 0pt}
     \setlength{\itemsep}{0pt} \settowidth
    {\labelwidth}{#1.}\sloppy}}{\end{list}}
\newcounter{itemlistc}
\newcounter{romanlistc}
\newcounter{alphlistc}
\newcounter{arabiclistc}

\newcommand{\fcaption}[1]{
        \refstepcounter{figure}
        \setbox\@tempboxa = \hbox{\footnotesize Fig.~\thefigure. #1}
        \ifdim \wd\@tempboxa > 6in
           {\begin{center}
        \parbox{6in}{\footnotesize\baselineskip=15pt Fig.~\thefigure. #1}
            \end{center}}
        \else
             {\begin{center}
             {\footnotesize Fig.~\thefigure. #1}
              \end{center}}
        \fi}
\newcommand{\tcaption}[1]{
        \refstepcounter{table}
        \setbox\@tempboxa = \hbox{\footnotesize Table~\thetable. #1}
        \ifdim \wd\@tempboxa > 6in
           {\begin{center}
        \parbox{6in}{\footnotesize\baselineskip=15pt Table~\thetable. #1}
            \end{center}}
        \else
             {\begin{center}
             {\footnotesize Table~\thetable. #1}
              \end{center}}
        \fi}
 1 
 1  
 1  
 1  
 1  
 1  
 1  

\font\ninerm=cmr9

\evensidemargin=\oddsidemargin  
\begin{document}
\begin{titlepage}
\begin{flushright}
{Extended Version}  \hfill hep-ph/9509278\\
{\small
DESY-96-029\\
TUIMP-TH-96/68\\
MSUHEP-60825}
\end{flushright}
\vskip 0.4in
\centerline{\normalsize\bf
PROBING ELECTROWEAK SYMMETRY BREAKING MECHANISM AT }  
\baselineskip=17pt
\centerline{\normalsize\bf             
THE LHC: A GUIDELINE FROM POWER COUNTING ANALYSIS }

\vspace*{1.2cm}
\centerline{\small {\bf  Hong-Jian He}~\footnote{
Electronic address: hjhe@desy.de} }
\baselineskip=17pt
\baselineskip=17pt
\centerline{\small\it
Theory Division, Deutsches Elektronen-Synchrotron DESY}
\centerline{\small\it D-22603 Hamburg, Germany}
\vspace*{0.6cm}
\centerline{\small {\bf Yu-Ping Kuang}~\footnote{
Electronic address: ypkuang@mail.tsinghua.edu.cn} }
\baselineskip=17pt
\centerline{\small\it
CCAST ( World Laboratory ), P.O.Box 8730, Beijing 100080, China }
\baselineskip=14pt
\centerline{\small\it
Institute of Modern Physics, Tsinghua University, Beijing 100084, China}
\vspace*{0.6cm}
\centerline{\small {\bf C.--P. Yuan}~\footnote{
Electronic address: yuan@pa.msu.edu }  }
\baselineskip=17pt
\centerline{\small\it
Department of Physics and Astronomy, Michigan State University } 
\baselineskip=14pt
\centerline{\small\it East Lansing, Michigan 48824, USA}
\baselineskip=18pt
\vspace{1cm}
\begin{abstract}
\noindent
We formulate the equivalence theorem as 
a theoretical criterion for sensitively
probing the electroweak symmetry breaking mechanism, and develop a 
precise power counting method for 
the chiral Lagrangian formulated electroweak theories. 
Armed with these, we perform a systematic analysis on the 
sensitivities of the scattering processes $W^\pm W^\pm \rightarrow 
W^\pm W^\pm$ and $q\bar{q}'\rightarrow W^\pm Z$ for testing all 
possible effective bosonic operators in the 
chiral Lagrangian formulated electroweak theories 
at the CERN Large Hadron Collider (LHC). The analysis shows that
these two kinds of processes are {\it complementary} in probing
the electroweak symmetry breaking sector.
\end{abstract}

\noindent
PACS number(s): 11.30.Qc, 11.15.Ex, 12.15.Ji, 14.70.-e
\vspace{0.3cm}
\begin{center}
( Version to be Published in {\it Mod. Phys. Lett. A} )
\end{center}

\end{titlepage}

\newpage
\setcounter{footnote}{00}
\renewcommand{\thefootnote}{\arabic{footnote}}
\renewcommand{\baselinestretch}{1.2}
\newpage

Recent LEP/SLC experiments  support the spontaneously broken 
$~SU(2)_L\otimes U(1)_Y~$ gauge theory as the correct 
description of electroweak interactions. 
However, a light standard model (SM) Higgs boson has 
not been found. The current experiments, allowing the Higgs mass
to range from $65$~GeV to about $O(1)$~TeV~\cite{higgs}, 
are not very sensitive to the electroweak symmetry 
breaking (EWSB) sector. Therefore, the EWSB mechanism 
remains a great mystery, and the probe of it has to include
both weakly and strongly interacting cases. If there is a
relatively light resonance originated from the EWSB mechanism, the probe 
would be easier. However, even if such a resonance is 
detected at the future colliders, 
it is still crucial to further test if it is associated with a
strong dynamics, because it is unknown {\it a priori} 
whether such a resonance trivially serves as the SM Higgs boson or comes 
from a more complicated mechanism~\cite{hill}. If the EWSB is driven by a 
strong dynamics with no new resonance much below the TeV scale,  
the probe becomes more difficult. In this paper, we study 
the latter case concerning the test 
at the CERN Large Hadron Collider (LHC).

The most economical description of the EWSB sector below the related new 
resonance scale is given by the electroweak chiral Lagrangian (EWCL) 
which can reflect both the heavy Higgs SM and 
other types of new strong dynamics. This general effective 
field theory approach is {\it complementary} to those specific model 
buildings. Following Ref.~\cite{ewcl,et4}, the EWCL can be formulated as
$$
\begin{array}{ll}
{\cal L}_{\rm eff} = {\cal L}_{\rm G} + {\cal L}_{\rm F}
+ {\cal L}^{(2)} + {\cal L}^{(2)\prime} 
+ \displaystyle\sum_{n=1}^{14} {\cal L}_n& 
= \displaystyle\sum_n \ell_n\displaystyle
\frac{f_\pi~^{r_n}}{\Lambda^{a_n}}
{\cal O}_n(W_{\mu\nu},B_{\mu\nu},D_\mu \!U,U,f,\bar{f}) ,
\end{array}
\eqno(1)                          
$$
where $~ {\cal L}_{\rm G} = -\frac{1}{4}W^a_{\mu\nu}W^{a\mu\nu}
                    -\frac{1}{4}B_{\mu\nu}B^{\mu\nu} ~$
and $~{\cal L}_{\rm F}~$ denotes the fermionic part. 
Here we concentrate on probing 
the new physics from all possible bosonic effective
operators so that we do not include the next-to-leading order 
fermionic operators in $~{\cal L}_{\rm F}~$. 
In (1), $~U=\exp [i\tau^a\pi^a /f_\pi ]~$ and $\pi^a$ is the
would-be Goldstone boson (GB) field.
$f_\pi=246$~GeV is the vacuum expectation value breaking the
electroweak gauge symmetry, and the effective cut-off
$~\Lambda \approx 4\pi f_\pi\approx 3.1~$TeV \cite{georgi} is 
the highest energy scale below which (1) is valid. 
The explicit expressions for nonlinear bosonic operators in
$~{\cal L}_{eff}~$ have been given by Refs.~\cite{ewcl,et4}, in 
which the leading order operator 
~${\cal L}^{(2)}=\frac{1}{4}f_\pi^2{\rm Tr}[(D^\mu U)(D_\mu U)^{\dagger}]$~ 
is universal, and the next-to-leading order 
operators ~${\cal L}^{(2)\prime}$,
~${\cal L}_{1\sim 11}$ ($CP$-conserving) and ~${\cal L}_{12\sim 14}$  
($CP$-violating) are model-dependent. 
Here, the dimensionless coefficients ~$\ell_n$~'s for these
next-to-leading order operators are related to the corresponding
notations ~$\alpha_n$~'s in Ref.~\cite{ewcl} by definition
$~~\alpha_n \equiv \left(\frac{f_\pi}{\Lambda}\right)^2\ell_n~~$.
From the consistency requirement of the chiral perturbation theory,
these coefficients ~$\ell_n$~'s can be naturally around of 
~$O(1)$~\cite{georgi}.

We know that only the longitudinal component $V^a_L$ 
of the weak-boson $V^a$ ($W^\pm$,$Z^0$), arising from ``eating'' the 
would-be Goldstone boson $\pi^a$ ($\pi^\pm ,\pi^0$), 
is sensitive to the EWSB sector, while 
the transverse component $V^a_T$ is not. For the strongly coupled
EWSB sector, the longitudinal $V_L$-scattering cross-sections are measurable 
at the LHC and  can thus probe the EWSB mechanism. 
The physical $V_L$-scattering amplitude is quantitatively related to 
the corresponding GB-amplitude by 
the electroweak Equivalence Theorem (ET)~\cite{et1}-\cite{et3}. In 
Ref.~\cite{et3}, we precisely formulate the ET as follows~\cite{et3} 
$$                                                                             
T[V^{a_1}_L,\cdots ,V^{a_n}_L;\Phi_{\alpha}]                                   
= C\cdot T[-i\pi^{a_1},\cdots ,-i\pi^{a_n};\Phi_{\alpha}]+ B ~~,
\eqno(2a)                                          
$$                                                                             
$$                                                                             
\begin{array}{l}                                                               
E_j \sim k_j  \gg  M_W , ~~~~~(~ j=1,2,\cdots ,n ~)~~,
\end{array}
\eqno(2b)                                         
$$
$$
\begin{array}{l}
C\cdot T[-i\pi^{a_1},\cdots ,-i\pi^{a_n};\Phi_{\alpha}]\gg B ~~,
\end{array}                                                                    
\eqno(2c)                                         
$$
where $~\pi^a$~'s  are GB fields, $\Phi_{\alpha}$ denotes other 
possible physical in/out states, $C\equiv C^{a_1}_{\rm mod}\cdots
C^{a_n}_{\rm mod}$ with 
$~C_{\rm mod}^a=1+O({\rm loop})~$ a renormalization scheme-dependent 
constant, ~$B\equiv\sum_{l=1}^n \\
(~C^{a_{l+1}}_{\rm mod}
\cdots C^{a_n}_{\rm mod}~T[v^{a_1},\cdots ,v^{a_l},-i\pi^{a_{l+1}},\cdots ,
-i\pi^{a_n};\Phi_{\alpha}] + {\rm permutations ~of}~v'
{\rm s ~and}~\pi '{\rm s}~)$ with $v^a\equiv v^{\mu}V^a_{\mu}$ and 
$v^{\mu}\equiv \epsilon^{\mu}_L-k^\mu /M_V = O(M_V/E),~(M_V=M_W,M_Z)$,
and $E_j$ is the energy of the $j$-th external line. 
The modification factor ~$C^a_{\rm mod}~$ has been generally studied in 
Refs.~\cite{et2}-\cite{et3}, and 
can be exactly simplified to unity in certain convenient renormalization 
schemes~\cite{et2',et3}. It is clear that the term $B$ in (2a) is 
$O(M_W/E)$-suppressed {\it relative} to the GB-amplitude 
~$C\cdot T[-i\pi^{a_1},\cdots]~$, 
but this does not mean that $B$ itself is necessarily of
$~O(M_W/E)$~ since the GB-amplitude contains positive powers
of $E$ in the CLEWT. In fact, our power counting rule [cf. (5)]
shows that, in the CLEWT, the leading term in $B$ for $V^a$-$V^b$
scatterings is of $~O(g^2)$ 
which is model-independent and of the same order as the leading pure 
$V_T$-amplitudes~\cite{et3}.
Hence $B$ {\it is insensitive to the EWSB mechanism}, and it serves
as an intrinsic background to {\it sensitively} probing 
the EWSB mechanism by the $V_L$-amplitude. 
Therefore, a sensitive probe at least requires
the GB-amplitude dominates
over $B$ to validate the equivalence between the $V_L$ and GB amplitudes 
in (2a). (2b)\footnote{
Condition (2b) is different from the usual condition $E\gg M_W$ 
for the total center of mass energy $E$.
An illustrating example is given in Ref.~\cite{et3}.}~ 
and (2c) are the precise conditions for this equivalence, 
and thus serve as the {\it necessary} conditions for 
sensitively probing the EWSB mechanism 
via $V_L$-scattering experiments.
Hence, we see the {\it profound physical content of the ET}:
it provides a necessary theoretical {\it criterion} 
for sensitively probing the EWSB mechanism,
and is much more than just a technical tool for 
simplifying explicit calculations.

To see the precise meaning of (2c), we consider
a certain perturbative expansion of the GB-amplitude. To a given order $N$ 
in the expansion, the amplitude $~T~$ can be written as 
~$T=\sum_{\ell=0}^NT_\ell$ with $T_0>T_1,\cdots, T_N$~. Let 
$~T_{\min}=\{T_0,\cdots,T_N \}_{\min}~$. Then, to the precision
of $T_{\min}~$, condition (2c) precisely implies 
$$
   T_{\min}[-i\pi^{a_1},\cdots ,-i\pi^{a_n};\Phi_{\alpha}]\gg B  ~~.
\eqno(3)                                                 
$$
For the CLEWT, the leading amplitude $~T_0~$ in (2a)
is of $~O(E^2)~$ and is model-independent. 
Thus, for distinguishing different strongly interacting EWSB
mechanisms, we have to consider the model-dependent next-to-leading order 
amplitude $T_1$ which can be of $~O(E^4)~$. Hence, 
we take $~T_{\min}=T_1~$ in (3). 
From (3), we can now theoretically define various levels
of the sensitivity for probing $T_1$ as follows. 
The probe is classified to
be {\it sensitive} if $T_1\gg B~$, 
{\it marginally sensitive} if $~T_1 >B~$
(but $~T_1\not\gg B~$), and {\it insensitive} if $~T_1\leq B~$.
 Note that in the following power counting analysis (cf. Table~1 and 2)
{\it both the GB-amplitude and the $B$-term are explicitly estimated by
our counting rule (5)}. The issue of numerically including/ignoring
$B$ in an explicit calculation is essentially {\it irrelevant} here.
If $~T_1\leq B~$, this means that the sensitivity is poor so that
the probe of $~T_1~$ is experimentally harder and requires a higher
experimental precision of at least $O(B)$ to test $~T_1~$.

To make a systematic global analysis on the sensitivity of each physical
scattering process for probing the new physics operators 
in the EWCL (1), we need a convenient method to obtain 
the scattering amplitudes contributed 
by all these operators. For this purpose, we generalize 
Weinberg's power counting rule for the ungauged nonlinear 
sigma model (NLSM)~\cite{wei} to the EWCL (1) and develop a precise 
power counting method for the CLEWT to {\it separately} count the 
power dependences on the energy $E$ and all relevant mass scales.
The original Weinberg's counting rule is to count the $E$-power
dependence ($D_E$) for a given $~L$-loop level $S$-matrix element 
$~T~$ in the NLSM. To generalize it to the EWCL, we further include 
the gauge boson, ghost and fermion fields and possible 
$~v_\mu$-factors  associated with external gauge-lines [cf.~(2a)]. 
From explicit derivations, we obtain the following counting formula for
the EWCL and in the high energy region 
$~\Lambda > E \gg M_W, m_t~$, 
$$ 
D_E = 2L+2+\displaystyle\sum_n {\cal V}_n
\left( d_n+\frac{1}{2}f_n -2\right) -e_v ~~,
\eqno(4)                       
$$
where ~${\cal V}_n$~ is equal to the number of type-$n$ vertices in ~$T$~,
~$d_n$ and $f_n$ are the numbers of derivatives and fermion-lines
at a type-$n$ vertex, respectively, and
$e_v$ is the number of possible external $v^\mu$-factors 
[cf. (2a) and below for the $B$-term].
Note that the counting rule (4) 
only holds for amplitudes {\it without any external $V_L$-line.}
Since there is non-trivial cancellation of the $E$-power factors
from the external $V_L$-polarizations in the $V_L$-amplitude
due to gauge-invariance, the $V_L$-amplitude cannot be directly 
counted by applying (4). However, there are no such $E$-power cancellations
on the RHS of (2a). Therefore (4) can be applied to 
amplitudes with external $V_L$-lines 
{\it by counting the RHS of the ET relation (2a)}.

Besides counting the power of $E$,
it is also crucial to {\it separately} count the
power dependences on the two typical mass scales in the EWCL, namely 
the vacuum expectation value $f_\pi$ and the effective cut-off 
$\Lambda$ , otherwise the result
will be off by orders of magnitudes since $~\Lambda /f_\pi 
\approx 4\pi > 12~$.
The $\Lambda$-dependence comes from two sources:~~ 
{\bf (i).} from tree vertices:
$T$ contains $~{\cal V}=\displaystyle\sum_n {\cal V}_n~$ vertices, 
each of which contributes a factor of $~1/\Lambda^{a_n}~$ [cf. (1)]
so that the total factor from ${\cal V}$-vertices is
$~1/\Lambda^{\sum_n a_n}~$;~ 
{\bf (ii).}  from loop-level: Since each loop brings in
a factor of $~(1/4\pi)^2 \approx (f_{\pi}/\Lambda )^2~$, the 
$~\Lambda$-dependence from $L$-loop 
contribution is $~~1/\Lambda^{2L}~~$.
Hence the total $~\Lambda$-dependence should be
$~~ 1/\Lambda^{\sum_na_n +2L}~~$. 
Let us denote the total dimension of ~$T$~ as ~$D_T~$, then
$~T~$ can always be
written as $~f_\pi^{D_T}~$ times some dimensionless function of 
$~E,~\Lambda,$ and $f_\pi$~ since the vacuum expectation value $~f_\pi~$
is generic to any spontaneously broken gauge theories. With these ready,
we can generally construct the following precise counting rule for 
$~T~$:\footnote{In (5) we still explicitly keep the loop factor
$~(1/4\pi )^{D_{EL}}~$ for generality, 
since the effective cut-off ~$\Lambda~$ denotes the lowest
new resonance scale and could be somehow lower than the 
theoretical upper bound
$~4\pi f_\pi\approx 3.1$~TeV for strongly coupled EWSB sector, 
as indicated by some model buildings.  For the case 
$~\Lambda \approx 4\pi f_\pi~$, this loop factor reduces to 
$~(f_\pi/\Lambda )^{D_{EL}}~$ as mentioned above.}
$$
\begin{array}{l}
T= c_T f_\pi^{D_T}\displaystyle 
\left(\frac{f_\pi}{\Lambda}\right)^{N_{\cal O}}
\left(\frac{E}{f_\pi}\right)^{D_{E0}}
\left(\frac{E}{4\pi f_\pi}\right)^{D_{EL}}
\left(\frac{M_W}{E}\right)^{e_v} H(\ln E/\mu) ~~,\\[0.5cm]
N_{\cal O}=\displaystyle\sum_n a_n~,~~~~ 
D_{E0}=2+\displaystyle\sum_n {\cal V}_n
\left(d_n+\frac{1}{2}f_n-2\right)~, ~~~~ 
D_{EL}=2L~,\\
\end{array}
\eqno(5)                                                  
$$ 
where the dimensionless coefficient $~c_T~$ contains 
possible powers of gauge 
couplings ($~g,e~$) and Yukawa 
couplings ($~y_f~$) from the vertices 
of $~T~$, which can be directly counted. 
$~H$ is a dimensionless function of $~\ln (E/\mu )~$ coming from 
loop corrections in the 
standard dimensional regularization~\cite{ewcl,dr}
(where $\mu$ is the relevant renormalization scale), 
and is thus insensitive to $E$. 
(Here we note that the dimensional regularization 
supplemented by the minimal subtraction
renormalization is particularly clean and convenient for effective theory
calculations, as emphasized in Ref.~\cite{dr}.)
Neglecting the insensitive factor $~H(\ln E/\mu)$, 
we can extract the main feature of
scattering amplitudes by simply applying (5) to
the corresponding Feynman diagrams.

Based upon the basic features of
the chiral perturbation expansion, we further build the following 
electroweak power counting hierarchy for the $S$-matrix elements, 
$$
\displaystyle{\frac{E^2}{f_\pi^2}\gg \frac{E^2}{f_\pi^2}
\frac{E^2}{\Lambda^2},~g\frac{E}{f_\pi}
\gg g\frac{E}{f_\pi}\frac{E^2}{\Lambda^2},~g^2\gg g^2\frac{E^2}{\Lambda^2}, 
~g^3\frac{f_\pi}{E}\gg g^3\frac{Ef_\pi}{\Lambda^2},
~g^4\frac{f_\pi^2}{E^2} \gg g^4\frac{f_\pi^2}{\Lambda^2}}~,
\eqno(6)                                               
$$
which, in the typical high energy region 
$E\in (750~{\rm GeV},~1.5~{\rm TeV})$ for instance, 
numerically gives (for $\Lambda \approx 4\pi f_\pi\approx 3.1$~TeV):
$$
\begin{array}{c}
(9.3,~37)\gg(0.55,~8.8),(2.0,~4.0)\gg(0.12,~0.93),(0.42,~0.42)\gg \\
(0.025,~0.099),(0.089,~0.045)\gg(5.3,~10.5)\times 10^{-3},
(19.0,~4.7)\times 10^{-3} \gg(1.1,~1.1)\times 10^{-3} ~.
\end{array}
\eqno(7)                                              
$$
The power counting hierarchy (6) provides a useful theoretical base for
our global classifications of various high energy scattering amplitudes.

In the literature (cf. Ref.~\cite{B-Y}), what usually done
is to study only a small subset of all effective operators in the
EWCL (1) for simplicity. But, to have a complete test of the EWSB sector
by distinguishing different kinds of dynamical models, 
it is necessary to know how to best measure all these operators 
through various high energy $VV$-fusion
and $q\bar{q}^{(\prime )}$-annihilation processes. 
For this purpose, our global power counting analysis provides a simple and
convenient way to quickly grasp the overall physical picture and
guides us to perform further elaborate numerical calculations.
In the following, we shall make the classifications for both the
$S$-matrix elements and the LHC event rates.

We first analyze the contributions of the fifteen effective
operators in (1) to all $V^a$-$V^b$ scatterings, which are 
dominated by the 4-GB-vertices [cf. the power counting rule (5)]. 
According to the hierarchy (6) and at the level of $S$-matrix elements,
Table~1 gives a complete sensitivity classification, 
which shows the {\it relevant} 
effective operators and the corresponding physical processes for 
probing the EWSB mechanism when calculating the scattering amplitude 
to the desired accuracy. In Table~1, 
MI and MD stand for model-independent and 
model-dependent operators, respectively. Here, for
simplicity we have taken $~\Lambda \approx 4\pi f_\pi ~$
whenever the one-loop MI contributions from ~${\cal L}^{(2)}~$ is concerned.
It is easy to change the one-loop factor back to $~(1/4\pi )^2~$ [cf.
(5)] when $~\Lambda < 4\pi f_\pi ~$. 
Also, we have explicitly estimated
all relevant contributions from the $B$-term. Here, $~B^{(i)}_{\ell}~$
($i=0,1,\cdots ; ~\ell =0,1\cdots$) denotes the $B$-term from the
~$\ell$-loop level $V_L$-amplitude containing $i$ external $V_T$-lines.
 From Table~1,
we first see that the MI operator $~{\cal L}^{(2)}~$ contained in 
~${\cal L}_{\rm MI}\equiv 
{\cal L}^{(2)}+{\cal L}_{\rm G}+{\cal L}_{\rm F}~$, 
mainly discriminating between the strongly and weakly 
interacting mechanisms, can be sensitively probed 
in the $~4V_L(\neq 4Z_L)~$ channel to the level of ~$O(E^2/f_\pi^2)$~. 
For the MD operators, the $4V_L$ channel can probe ~${\cal L}_{4,5}$~ 
most sensitively. The contributions of ~${\cal L}^{(2)\prime}$~ and 
${\cal L}_{2,3,9}$ to this channel lose the $E$-power dependence by a 
factor of $2$. Hence this channel is less sensitive to these operators.
The ~$4V_L$~ channel cannot probe ~${\cal L}_{1,8,11\sim 14}~$ (which
can only be probed via channels with $V_T$~'s).
Among $~{\cal L}_{1,8,11\sim 14}~$, 
the contributions from $~{\cal L}_{11,12}~$ to channels 
with $V_T$('s) are most important though they 
are still suppressed by a factor of $~gf_\pi /E$~ relative to the
leading contributions from $~{\cal L}_{4,5}~$ to the ~$4V_L$~ channel.
~${\cal L}_{1,8,13,14}$~ are generally suppressed by higher powers of
~$gf_\pi /E$~ and are thus the least sensitive.

Table~2 classifies all $q\bar{q}^{(\prime )}$-annihilation processes.
The operator $~{\cal L}_{\rm MI}~$ can be probed via tree-level constant
$~O(g^2)~$ amplitude through either ~$V_LV_L$ ~or ~$V_TV_T$~ final states,
which are not enhanced by high energy $E$-powers.
Among all next-to-leading order operators, 
the probe of $~{\cal L}_{2,3,9}~$ is most sensitive via 
$~q\bar{q}^{(\prime)}\rightarrow W^+_LW^-_L$~ amplitude
 and the probe of $~{\cal L}_{3,11,12}~$ is best via
$~q\bar{q}^{\prime}\rightarrow W^\pm_LZ_L~$ amplitude, to the precision of
$~g^2{E^2\over\Lambda^2}~$. For operators 
$~{\cal L}_{1,8;13,14}~$, the largest amplitudes are
$~T_1[q\bar{q};W^+_LW^-_T/W^+_TW^-_L]~$ and
$~T_1[q\bar{q}^{\prime};W^\pm_LZ_T/W^\pm_TZ_L]~$, which are
at most of $~O\left(g^3\frac{Ef_\pi}{\Lambda^2}\right)~$.
The contributions to total cross sections 
from above amplitudes can exceed that from the 
corresponding $~B= O\left(g^2\frac{M_W^2}{E^2}\right)~({\rm for}~
V_LV_L)~{\rm or}~O\left(g^2\frac{M_W}{E}\right)~ 
({\rm for}~V_LV_T)~$ in the high energy
region when polarizations are summed up. 
The next-to-leading order operators $~{\cal L}_{4,5,6,7,10}~$ do not
contribute to $q\bar{q}^{(\prime )}$-annihilations at the $~1/\Lambda^2$-order
and thus will be best probed via $VV$-fusions (cf. Table~1).

Before further classifying the various contributions to a given process
at the event rate level for the LHC, we have compared some of our results 
with those available in Ref.~\cite{bdv} from precise calculations,
to test the above power counting method. The authors of
Ref.~\cite{bdv} and we have used the effective-$W$ approximation 
(EWA)~\cite{w,mike-new} for computing the event rates. 
Two typical processes for $WW$-fusion and $q\bar{q}$-annihilation 
are compared in Fig.~1a and 1b, respectively. The event rates
$~R_{\alpha\beta\gamma\delta (\ell )}~$ and 
$~R_{\alpha\beta (\ell )}~$ are calculated  up to one-loop level
for the two processes. 
(Here $~\alpha ,\beta ,\gamma ,\delta = L,T~$ specify the polarizations
of the incoming/out-going $W^\pm$ or $Z^0$ gauge bosons, 
and $~\ell =0,1~$ denote the
tree and one-loop level contributions, respectively.)
The comparison in Fig.~1 shows that 
the agreements are within a factor of $~2~$ or even better. 
So, our simple power counting rule (5) does conveniently give reasonable 
systematic estimates and is thus useful for making global analyses on 
probing the EWSB mechanisms at the LHC and future linear colliders.

Then, we calculate the number of events 
per $[100\,{\rm fb}^{-1}~{\rm GeV}]$ 
contributed from each next-to-leading order effective operator 
at the LHC by the power counting rule (5) combined 
with the EWA.\footnote{
We clarify that the theoretical criterion (3) is {\it necessary}
but not sufficient at the event rate level, since the leading 
$B$-term, as an intrinsic background to any strong $V_L$-$V_L$ scattering 
process, denotes a universal part of the full backgrounds~\cite{mike,et3}.
The sufficiency will of course require detailed numerical analyses on 
the detection efficiency for suppressing 
the full backgrounds to observe the specific decay
mode of the final state (as discussed in Ref.~\cite{B-Y}).
This is beyond our present first step global analysis 
and will be left to a future detailed numerical study with
this work as a useful guideline.}~ 
~In the following numerical analysis we typically take 
$~\Lambda \approx 4\pi f_\pi \approx 3.1~$TeV. But we keep in mind that
our estimates for the number of events contributed by
the next-to-leading order operators will be increased by a factor
of $~(3.1~{\rm TeV}/\Lambda)^2~$ for $~\Lambda < 3.1~$TeV 
in the energy region below  $~\Lambda~$.
We first consider the $W^+W^+$
channel which is most important for the 
non-resonance scenario~\cite{mike,B-Y}.
In Fig.~2, the event rate $~|R_1|~$ contributed 
from each next-to-leading order operator 
is separately shown for the $W^+W^+$ channel for $\ell_n \simeq O(1)$ 
with the polarizations of the initial and final states summed over.
We note that
the experiments actually contain the contributions from {\it all} 
operators and are thus more complicated. 
For simplicity and clearness, one can make
the well-known naturalness assumption (i.e., contributions from different 
operators do not accidentally cancel each other), as widely adopted
in the literature~\cite{d-v}, and estimate the bounds on
each single operator.  We can clearly see, from Fig.~2, 
the sensitivities for probing these new physics 
operators  by comparing the event rates ($|R_1|$) contributed from these 
operators with the rate ($|R_B|$) from the $B$-term 
(which serves as a {\it necessary} criterion as defined above). 
In Fig.~2a, the event rates from ~${\cal L}_{4,5}~$ are larger than
that from the $B$-term when $~E> 600~$GeV, while those from
~${\cal L}_{3,9,11,12}~$ can exceed the rate from $B$ only if
$~E> 860~$GeV. In Fig.~2b, 
the rates from $~{\cal L}^{(2)\prime}~$ and
~${\cal L}_{1,2,8,13,14}$~ are all 
below the rate from $B$ for a wide range of energy up to about $2~$TeV. 
 We thus conclude that for coefficients $~\ell_n \simeq O(1)~$,
 the probe of ~${\cal L}_{4,5}~$ is most 
sensitive, that of $~{\cal L}_{3,9,11,12}~$ is marginally sensitive, 
and that of $~{\cal L}^{(2)\prime}, {\cal L}_{1,2,8,13,14}$~ is 
insensitive. In this case, a precise test of the marginal operators
$~{\cal L}_{3,9,11,12}~$ via the $W^+ W^+$ channel
 requires including the $B$-term in calculating the
weak-boson scattering amplitudes which have to be obtained from 
a full calculation beyond the EWA~\footnote{
It is well-known that the validity of the EWA
requires $~M_{VV}\gg 2M_W~$ and the scattering angles big enough to be
away from kinematic  singularities~\cite{w,mike-new}. The condition
$~M_{VV}\gg 2M_W~$ shows that the EWA
has a precision similar to that of the ET. 
The $V_T$-$V_L$ interference term ignored in the usual EWA~\cite{w} 
is also of the same order as the $B$-term in (2a)~\cite{w}.
The sole purpose of a recent paper (hep-ph/9502309) by A. Dobado et al
was to avoid the ET, but still within the usual EWA, for increasing the 
calculation precision and extending the results to lower energy regions.
This approach is, however, inconsistent because both the ET and EWA are 
valid only in the {\it high energy regime} and have 
{\it similar precisions} as explained above.}.~
It also implies that a higher luminosity of the collider is 
needed for probing these operators via the $W^+W^+$ 
productions.
For the case with $~\ell_n \simeq O(5\sim 10)~$, 
the probe of $~{\cal L}_{3,9,11,12}~$ 
can become sensitive, while
~${\cal L}^{(2)\prime}$~ and ~${\cal L}_{1,2,8,13,14}$~ still cannot be
sensitively probed in the $W^+ W^+$ channel. A similar conclusion
holds for $W^-W^-$ channel except that its event rate is lower by
about a factor of ~$3\sim 5$~ in the TeV region since the quark luminosity
for producing $W^-W^-$  pairs is smaller than that for $W^+W^+$ pairs
in pp collisions.

Next we compute the event rates for the important 
~$q\bar{q}'\rightarrow W^+ Z~$ process.  Fig.~3
shows that for $~\ell_n\simeq O(1)$~, the probe of 
~${\cal L}_{3,11,12}~$ are sensitive when $~E>750~$GeV, 
while that of ~${\cal L}_{8,9,14}$~ are 
marginally sensitive when $~E>950~$GeV. The probe of 
~${\cal L}^{(2)^\prime}$~ becomes marginally
sensitive when ~$E>1.4$~TeV, 
and that of ~${\cal L}_{1,2,13}~$ is insensitive for $~E<1.9$~TeV. 
(We note that ~${\cal L}_1~$ and ~${\cal L}^{(2)^\prime}$~ can 
be better measured at the low energy experiments through $S$ and $T$ 
parameters, respectively, while ~${\cal L}_{13,14}~$ can be more
sensitively probed via $~e^-\gamma\rightarrow \nu_e W^-_LZ^0_L,
e^-W^-_LW^+_L~$ processes at the future TeV linear collider~\cite{LC}.)  
The event rate for ~$q\bar{q}'\rightarrow W^+ Z~$
 is slightly higher than that of ~$q\bar{q}'\rightarrow W^- Z~$
by about a factor of ~$1.5$~(or smaller) 
due to the higher quark luminosity for 
producing $W^+$ bosons in pp collisions. 
Hence, the similar conclusion holds
for the ~$q\bar{q}'\rightarrow W^- Z~$ process. 
Comparing the above $W^\pm W^\pm\rightarrow W^\pm W^\pm$ fusions 
and $q\bar{q}'\rightarrow W^\pm Z$ annihilations,
we see that these two kind of processes are 
{\it complementary} to each other in probing 
the effective operators of the EWCL (1).

In summary, the analyses presented in this paper
are consistently performed based upon 
the electroweak power counting rule (5)
combined with the effective-$W$ method.
In Table~1 and 2, the sensitivity classifications  
are summarized at the level of the
$S$-matrix elements and for all $~V^aV^b\rightarrow V^cV^d~$
and $~q\bar{q}^{(\prime )}\rightarrow V^aV^b~$
processes according to the electroweak power counting hierarchy (6).
Estimates\footnote{cf. footnote-$3$.}~ 
on the event rates at the $14$~TeV LHC
with an integrated luminosity of $~100~{\rm fb}^{-1}~$ are 
given for both $~W^+W^+\rightarrow W^+W^+$~ (cf. Fig.~2) and
~$q\bar{q}'\rightarrow W^+ Z^0$~ (cf. Fig.~3) channels, which are shown to be
{\it complementary}  in probing the operators in (1).
By these, we give a clear 
physical picture for globally classifying all 
bosonic effective operators to probing the underlying EWSB mechanism.
This provides a useful guideline for 
future detailed numerical computations and analyses. 
The extension of our analysis to future linear
colliders are given in Ref.~\cite{LC}.

\vspace{0.8cm}
\noindent
{\bf Acknowledgements}~~~~
We thank
Sally Dawson for useful conversations on the EWA used in Ref.~\cite{bdv}.
We are also grateful to 
Mike Chanowitz, Tao Han, and Peter Zerwas for useful discussions 
and carefully reading the
manuscript.  H.J.H. is supported by the AvH of Germany and the U.S. DOE; 
Y.P.K. is supported by the National NSF of China 
and the FRF of Tsinghua Univ.; C.P.Y. is supported in part by U.S. NSF. 


\vspace{1.2cm}

\vspace{1.5cm}
\noindent
{\bf Table Captions}

\noindent
{\bf Table~1.} Global classification for probing direct and indirect EWSB 
information at the level of $S$-matrix elements (A). $^{(a)}$\\
Notes:\\ {\footnotesize
$^{(a)}$ The contributions from
${\cal L}_{1,2,13}$ are {\it always} associated 
with a factor of $\sin^2\theta_W$, unless specified otherwise. 
Also, for contributions to the $B$-term in 
a given $V_L$-amplitude, we list them
separately with the $B$-term specified.\\ 
$^{(b)}$ MI $=$ model-independent, MD $=$ model-dependent.\\
$^{(c)}$ There is no contribution when all the external lines are 
electrically neutral.\\
$^{(d)}$ $B_0^{(1)}\simeq T_0[2\pi ,v,V_T]~(\neq T_0[2\pi^0 ,v^0,Z_T])$,~
$B_0^{(3)}\simeq T_0[v,3V_T]~(\neq T_0[v^0,3Z_T])$.\\
$^{(e)}$  $T_1[2V_L,2V_T]=T_1[2Z_L,2W_T],~T_1[2W_L,2Z_T]$, ~or
$~T_1[Z_L,W_L,Z_T,W_T]$.\\
$^{(f)}$ ${\cal L}_2$ only contributes to $T_1[2\pi^\pm ,\pi^0,v^0]$ and
$T_1[2\pi^0,\pi^\pm ,v^\pm ]$ at this order; 
${\cal L}_{6,7}$ do not contribute
to $T_1[3\pi^\pm ,v^\pm ]$.\\
$^{(g)}$  ${\cal L}_{10}$ contributes only 
to $T_1[\cdots ]$ with all the external
lines being electrically neutral.\\
$^{(h)}$ $B_0^{(2)}$ is dominated by $~T_0[2V_T,2v]~$ since 
$~T_0[\pi,2V_T,v]~$ contains a suppressing factor $\sin^2\theta_W$ 
as can be deduced from $~T_0[\pi,3V_T]~$ 
times the factor $~v^\mu =O\left(\frac{M_W}{E}\right)~$.\\
$^{(i)}$ Here, $T_1[2W_L,2W_T]$ contains a coupling 
$e^4=g^4\sin^4\theta_W$.\\
$^{(j)}$ ${\cal L}_2$ only contributes to $T_1[3\pi^\pm ,v^\pm ]$.\\
$^{(k)}$ ${\cal L}_{1,13}$ do not contribute to 
$T_1[2\pi^\pm ,2v^\pm ]$.  }

\noindent
{\bf Table~2.} 
Global classification for probing direct and indirect EWSB
information at the level of $S$-matrix elements (B). $^{(a)}$

\vspace{1.0cm}
\noindent
{\bf Figure Captions}

\noindent
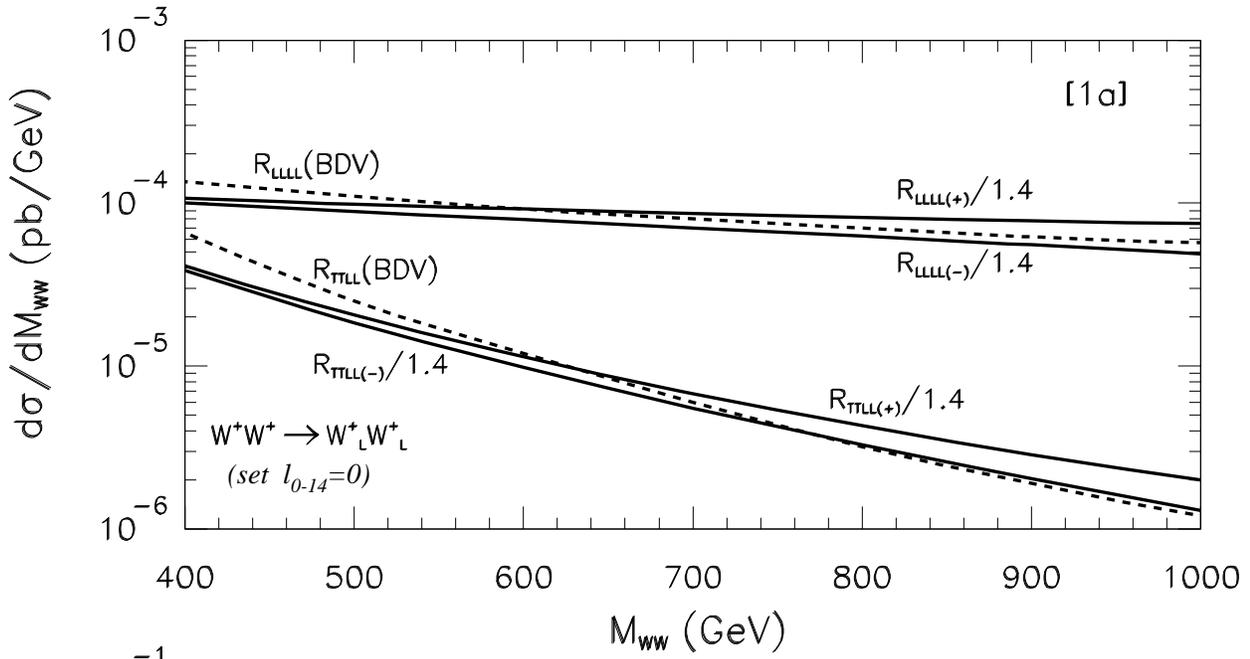
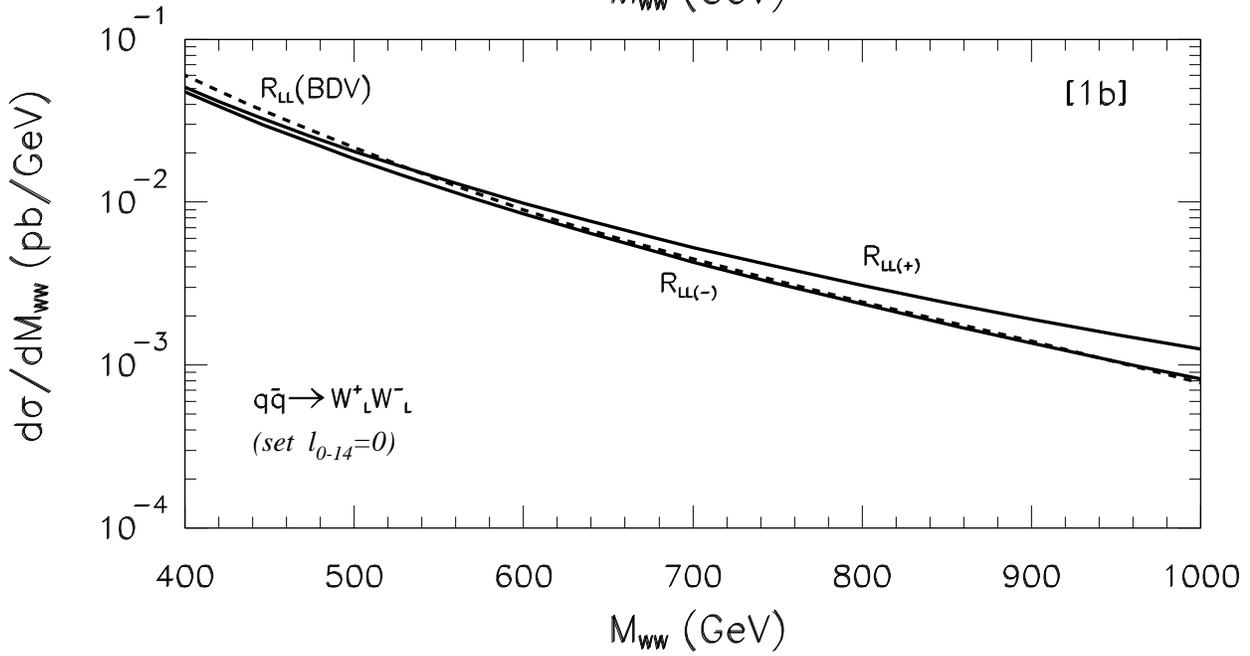
{\bf Fig.~1.} Comparsion with the calculations of Ref.~\cite{bdv}
up to $1$-loop for $\sqrt{S}=40$~TeV. 
The solid and dashed lines are given by
our power counting analysis and Ref.~\cite{bdv}, respectively. 
($~R_{\alpha\beta\gamma\delta (\pm )}
= R_{\alpha\beta\gamma\delta (0)}\pm |R_{\alpha\beta\gamma\delta (1)}|~$
and 
$~R_{\alpha\beta (\pm )}
=R_{\alpha\beta (0)}\pm |R_{\alpha\beta (1)}|~$, where
$~R_{\alpha\beta\gamma\delta (\ell )}~$ and
$~R_{\alpha\beta (\ell )}~$ are explained in the text.)

\noindent
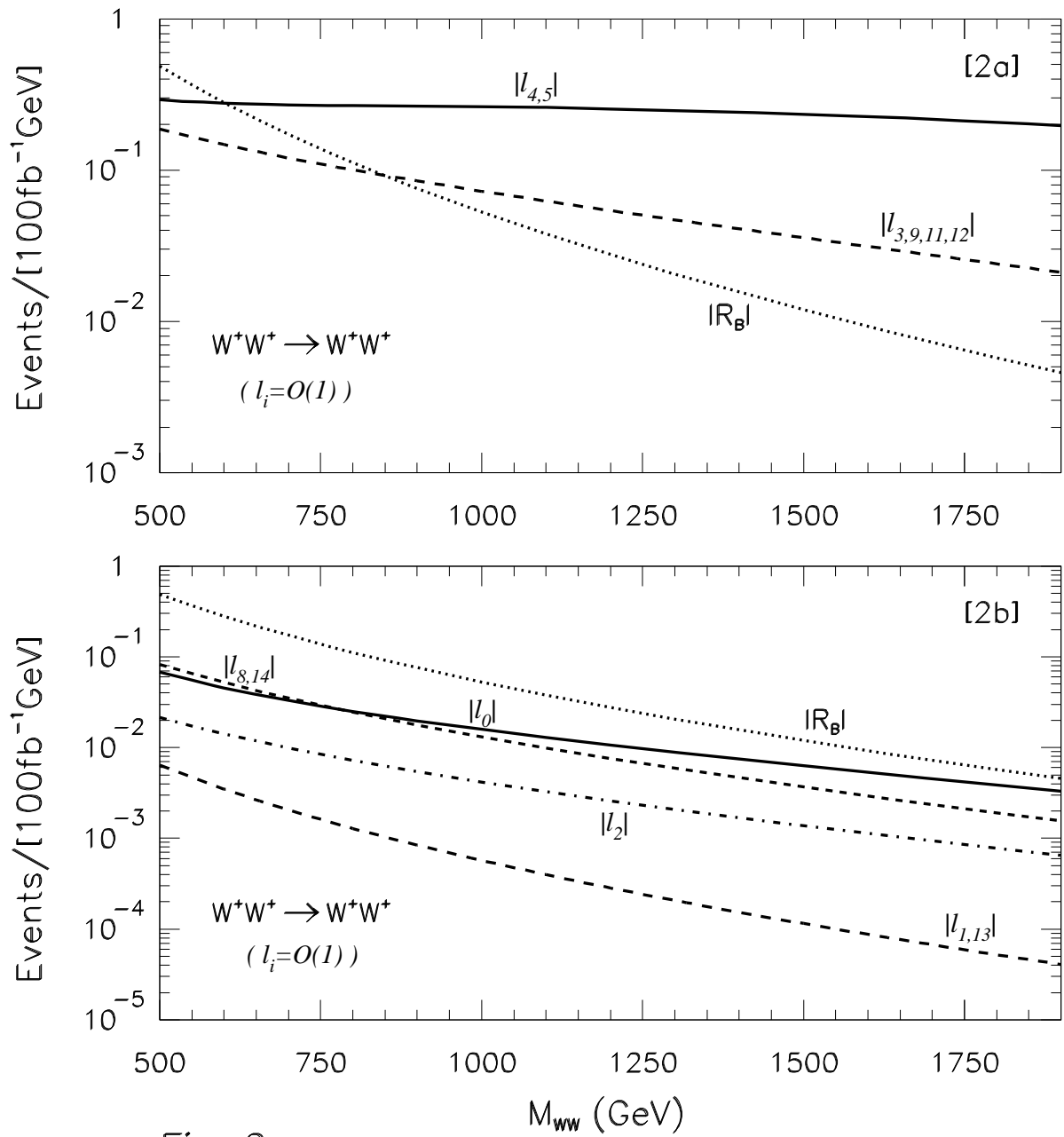
{\bf Fig.~2.} Sensitivities of probing ${\cal L}^{(2)\prime}$ 
and ${\cal L}_{1\sim 14}$ 
in the $W^+W^+$ channel at the $14$~TeV LHC.

\noindent
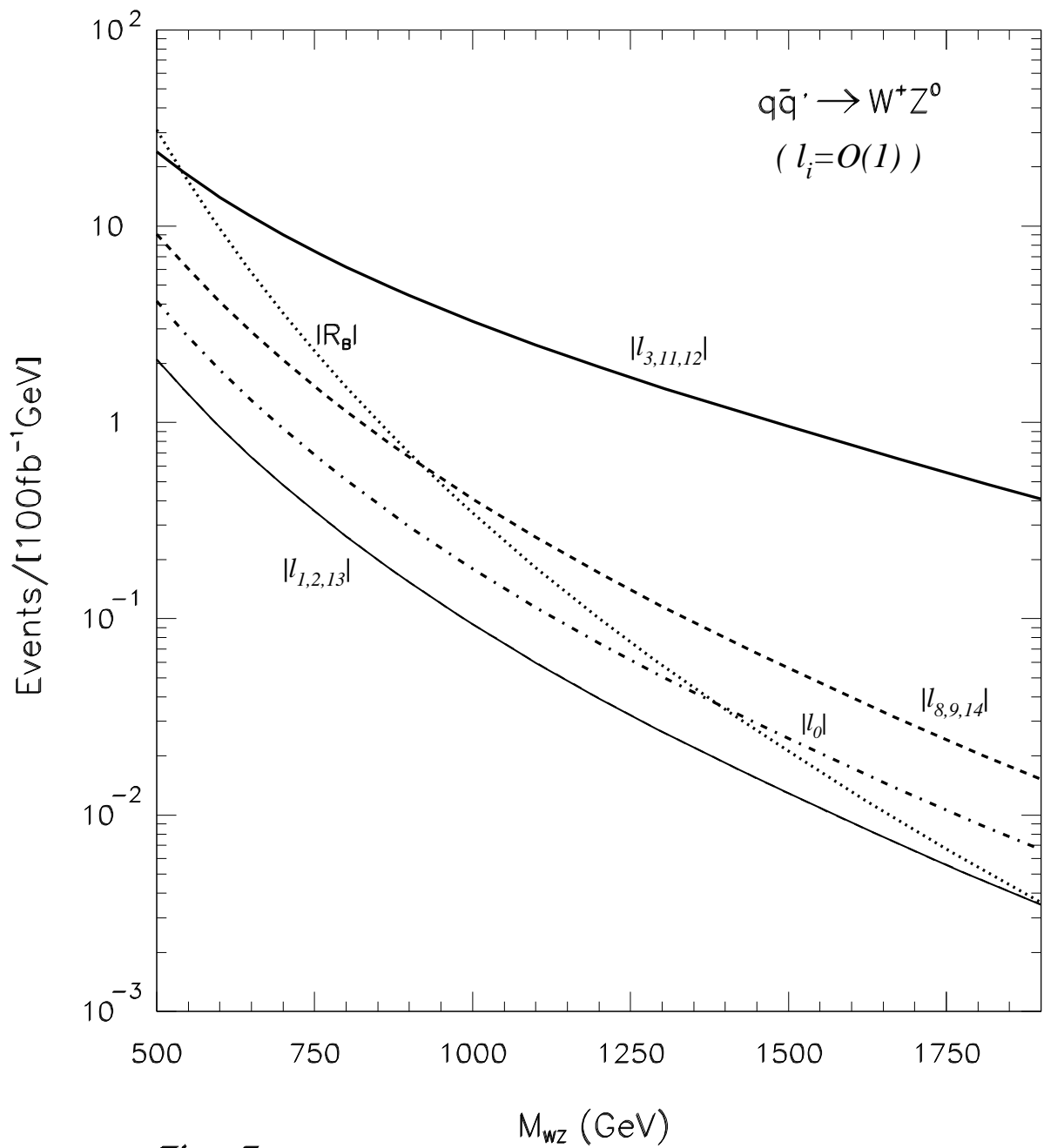
{\bf Fig.~3.} Sensitivities of probing ${\cal L}^{(2)\prime}$ 
and ${\cal L}_{1\sim 14}$ 
in the $~q\bar{q}'\rightarrow W^+ Z^0~$ channel at the $14$~TeV LHC.

\newpage

\tabcolsep 1pt
\renewcommand{\baselinestretch}{1}


\begin{table}[t]  
\begin{center}

{\bf Table~1.} 
Global classification for probing direct and indirect\\ 
EWSB information at the level of $S$-matrix elements (A). $^{(a)}$

\vspace{0.2cm}


\small

\begin{tabular}{||c||c|c|c||} 
\hline\hline
& & &  \\
Required Precision
&  Relevant Operators
&  Relevant Amplitudes
&  MI or MD $^{(b)}$ \\
& & & ? \\
\hline\hline
   $O\left(\frac{E^2}{f_{\pi}^2}\right)$ 
&  ${\cal L}_{\rm MI}~(\equiv 
   {\cal L}_{\rm G}+{\cal L}_{\rm F}+{\cal L}^{(2)}) $
&  $ T_0[4V_L] (\neq T_0[4Z_L]) $
&  MI \\
\hline
\parbox[t]{2.2cm}{ 
~~\\
~~\\
$O\left(\frac{E^2}{f_\pi^2}
 \frac{E^2}{\Lambda^2},~g\frac{E}{f_\pi}\right)$\\
~~\\
~~ }
&  \parbox[t]{2.8cm}{ 
  ${\cal L}_{4,5}$\\
  ${\cal L}_{6,7}$\\
  ${\cal L}_{10}$\\
  ${\cal L}_{\rm MI}$\\
  ${\cal L}_{\rm MI}$ }
&  \parbox[t]{5.0cm}{ 
  $T_1[4V_L]$\\
  $T_1[2Z_L,2W_L],~T_1[4Z_L]$\\
  $T_1[4Z_L]$\\
  $T_0[3V_L,V_T] ~(\neq T_0[3Z_L,Z_T])$\\
  $T_1[4V_L]$\\ }
&  \parbox[t]{0.8cm}{ 
  MD\\
  MD\\
  MD\\
  MI\\
  MI }\\
\hline
\parbox[t]{2.2cm}{ 
~~\\
~~\\
$O\left(g\frac{E}{f_\pi}\frac{E^2}{\Lambda^2},~g^2\right)$\\
~~\\
~~\\
~~  }
& \parbox[t]{2.8cm}{ 
  $ {\cal L}_{3,4,5,9,11,12} $\\
  $ {\cal L}_{2,3,4,5,6,7,9,11,12} $\\
  $ {\cal L}_{3,4,5,6,7,10} $\\
  $ {\cal L}_{\rm MI} $\\ 
  $ {\cal L}_{\rm MI} $\\
  $ {\cal L}_{\rm MI} $} 
&  \parbox[t]{5.8cm}{ 
  $T_1[3W_L,W_T]$\\
  $T_1[2W_L,Z_L,Z_T],~T_1[2Z_L,W_L,W_T]$\\
  $T_1[3Z_L,Z_T]$\\
  $T_0[2V_L,2V_T],~T_0[4V_T]~~^{(c)}$\\
  $T_1[3V_L,V_T]$\\
  $B^{(0)}_0 \simeq T_0[3\pi ,v]~(\neq T_0[3\pi^0 ,v^0])$  }
&   \parbox[t]{0.8cm}{ 
  MD\\
  MD\\
  MD\\
  MI\\
  MI\\
  MI }\\
\hline
  $O\left(\frac{E^2}{\Lambda^2}\right)$
& ${\cal L}^{(2)\prime}$ 
& $T_1[4W_L],~T_1[2W_L,2Z_L]$
& MD \\
\hline
   \parbox[t]{2.5cm}{
        ~~\\
~~\\
~~\\
$O\left(g^2\frac{E^2}{\Lambda^2},~g^3\frac{f_\pi}{E}\right)$ \\ 
~~\\          
~~\\
~~ }
&  \parbox[t]{3.0cm}{ 
   ${\cal L}_{\rm MI}$\\
   ${\cal L}_{2,3,9}$\\
   ${\cal L}_{3,11,12}$\\
   ${\cal L}_{2,3,4,5,8,9,11,12,14}$\\
   ${\cal L}_{1\sim 9,11\sim 14}$ \\
   ${\cal L}_{4,5,6,7,10}$ \\
   ${\cal L}_{{\rm MI},2,3,4,5,6,7,9\sim 12}$ }
&  \parbox[t]{6.3cm}{  
   $T_0[V_L,3V_T],T_1[2V_L,2V_T], B^{(1,3)}_0~~^{(c,d)} $\\
   $T_1[4W_L]$\\
   $T_1[2Z_L,2W_L]$\\
   $T_1[2W_L,2W_T]$\\
   $T_1[2V_L,2V_T]~~~^{(e)}$\\
   $T_1[2Z_L,2Z_T]$\\
{\footnotesize  $B^{(0)}_1 \simeq T_1[3\pi ,v]~~^{(f,g)}$}  }
&  \parbox[t]{2.0cm}{ 
  MI\\
  MD\\
  MD\\
  MD\\
  MD\\
  MD\\
  MI $+$ MD\\  }\\
\hline
\parbox[t]{2.75cm}{
~~\\
~~\\
  $O\left(g^3\frac{Ef_\pi}{\Lambda^2},~g^4\frac{f_\pi^2}{E^2}\right)$\\
~~\\
~~\\}
& \parbox[t]{3.3cm}{ 
${\cal L}_{{\rm MI},1,2,3,8,9,11\sim 14}$ \\
${\cal L}_{4,5}$\\  
${\cal L}_{6,7,10}$ \\
${\cal L}_{2\sim 5,8,9,11,12,14}$\\
${\cal L}_{\rm MI}$   }
& \parbox[t]{5.5cm}{ 
$T_1[V_L,3V_T]~(\neq T_1[Z_L,3Z_T])$\\
$T_1[V_L,3V_T]$\\
$T_1[V_L,3V_T]~(\neq T_1[W_L,3W_T])~~^{(g)}$ \\
{\footnotesize $B^{(1)}_1\simeq T_1[2\pi ,V_T,v]$} \\
{\footnotesize $B^{(2)}_0\simeq T_0[2V_T,2v]$}$~~^{(c,h)}$  }
& \parbox[t]{2.0cm}{ MI$+$MD \\
                     MD\\ 
                     MD\\
                     MD\\
                     MI }\\
\hline
\parbox[t]{2.0cm}{ 
~~\\
~~\\
~~\\
~~\\
~~\\
  $O\left((g^2,g^4)\frac{f_\pi^2}{\Lambda^2}\right)$\\
~~\\
~~\\
~~\\
~~\\
~~}
& \parbox[t]{3.1cm}{ 
  ${\cal L}^{(2)\prime} $ \\
  ${\cal L}_1$\\
  ${\cal L}_{{\rm MI},1\sim 5,8,9,11\sim 14}$\\
  ${\cal L}_{{\rm MI},1\sim 9,11\sim 14}$\\
  ${\cal L}_{{\rm MI},1,4,5,6,7,10}$\\
  ${\cal L}_{1,2,8,13,14}$\\
  ${\cal L}_{{\rm MI},1\sim 9,11\sim 14}$\\
  ${\cal L}_{{\rm MI},4,5,6,7,10}$\\
  ${\cal L}_{{\rm MI},1\sim 5,8,9,11\sim 14}$\\
  ${\cal L}_{{\rm MI},1\sim 9,11\sim 14}$\\
  ${\cal L}_{{\rm MI},4,5,6,7,10}$ } 
& 
\parbox[t]{6.4cm}{
$T_1[2V_L,2V_T],B^{(0)}_1\simeq T_1[3\pi ,v]~~^{(c)}$ \\
  $ T_1[2W_L,2W_T]~~^{(i)} $\\
  $ T_1[4W_T] $\\
  $ T_1[4V_T]~(\neq T_1[4W_T],T_1[4Z_T]) $\\
  $ T_1[4Z_T] $\\
{\footnotesize  $ B_1^{(0)}\simeq T_1[3\pi ,v]~~^{(c,j)} $\\
  $ B_1^{(0)}\simeq T_1[2\pi ,2v]~~^{(c,k)} $\\
  $ B_1^{(0)}\simeq 
                      T_1[2\pi ,2v](\neq T_1[2\pi^\pm ,2v^\pm ])~^{(g)} $ \\
  $ B_1^{(2)}\simeq T_1[\pi^\pm ,2W_T,v^\pm ] $\\
  $ B_1^{(2)}\neq T_1[\pi^\pm ,2W_T,v^\pm ],
     T_1[\pi^0 ,2Z_T,v^0] $\\
  $ B_1^{(2)}\simeq T_1[\pi^0 ,2Z_T,v^0] $ } }
&  
\parbox[t]{2.0cm}{ 
  MD\\
  MD\\
  MI$+$MD\\
  MI$+$MD\\
  MI$+$MD\\
  MD\\
  MI$+$MD\\
  MI$+$MD\\
  MI$+$MD\\
  MI$+$MD\\
  MI$+$MD  }\\
& & &  \\
\hline\hline 
\end{tabular}
\end{center}
\end{table}


\newpage

\tabcolsep 1pt
\begin{table}[t]  
\begin{center}

{\bf Table~2.} 
Global classification for probing direct and indirect\\ 
EWSB information at the level of $S$-matrix elements (B). 
$^{(a)}$

\vspace{0.5cm}


\small

\begin{tabular}{||c||c|c|c||} 
\hline\hline
& & &  \\
~Required Precision~
&  Relevant Operators
&  Relevant Amplitudes
&  MI or MD $^{(b)}$ \\
& & & ? \\
\hline\hline
& & & \\
     $O(g^2)$  
&    $~{\cal L}_{\rm MI}~(\equiv 
       {\cal L}_{\rm G}+{\cal L}_{\rm F}+{\cal L}^{(2)}) ~$
& $T_0[q\bar{q};V_LV_L],~ T_0[q\bar{q};V_TV_T]$
& MI \\
& & & \\
\hline
& & & \\
   \parbox[t]{2.5cm}{
        ~~\\
        ~~\\
$O\left(g^2\frac{E^2}{\Lambda^2},~g^3\frac{f_\pi}{E}\right)$ \\ 
        ~~\\          
        ~~ }
&  \parbox[t]{3.0cm}{ 
   ${\cal L}_{2,3,9}$\\
   ${\cal L}_{3,11,12}$\\
   ${\cal L}_{\rm MI}$\\ 
   ${\cal L}_{\rm MI}$\\
   ${\cal L}_{\rm MI}$ }
&  \parbox[t]{4.0cm}{  
   $T_1[q\bar{q};W_LW_L] $\\
   $T_1[q\bar{q};W_LZ_L] $ \\
   $T_0[q\bar{q};V_LV_T] $\\
   $T_1[q\bar{q};V_LV_L] $\\
 {\footnotesize  $B_0^{(1)}$}$\simeq T_0[q\bar{q};V_T,v]$ }
&  \parbox[t]{0.8cm}{ 
  MD\\
  MD\\
  MI\\
  MI\\
  MI }\\
& & & \\
\hline
& & & \\
\parbox[t]{2.5cm}{
~~\\
  $O\left(g^3\frac{Ef_\pi}{\Lambda^2},
   ~g^4\frac{f_\pi^2}{E^2}\right)$ \\
~~  }
& \parbox[t]{3.3cm}{ 
${\cal L}_{1,2,3,8,9,11\sim 14}$\\
${\cal L}_{\rm MI}$ \\
${\cal L}_{\rm MI}$  }
& \parbox[t]{4.0cm}{ 
$T_1[q\bar{q};V_LV_T]$\\  
$T_1[q\bar{q};V_LV_T]$\\
{\footnotesize $B_0^{(0)}$}$\simeq T_0[q\bar{q};\pi ,v]~^{(c)}$  }
& \parbox[t]{0.8cm}{ MD \\
                     MI \\
                     MI }\\
& & & \\
\hline
& & & \\
\parbox[t]{2.0cm}{ 
~~\\
  $O\left((g^2,g^4)\frac{f_\pi^2}{\Lambda^2}\right)$\\
~~}
& \parbox[t]{3.1cm}{ 
  ${\cal L}^{(2)\prime} $ \\
  ${\cal L}_{1,2,3,8,9,11\sim 14}$\\
  ${\cal L}_{\rm MI}$ } 
& 
\parbox[t]{5.4cm}{$ T_1[q\bar{q};V_LV_L]$\\  
                  $ T_1[q\bar{q};V_TV_T]$,~
 {\footnotesize $B_1^{(0)}$}$\simeq T_1[q\bar{q};\pi ,v]$ \\  
  $ T_1[q\bar{q};V_TV_T]$,~
 {\footnotesize $B_1^{(0)}$}$\simeq T_1[q\bar{q};\pi ,v]$   }
&  \parbox[t]{0.8cm}{ 
  MD\\
  MD\\
  MI  }\\
& & &  \\
\hline\hline 
\end{tabular}
\end{center}
\end{table}
\begin{table}[t]
\begin{center}

\vspace{0.4cm}
\begin{tabular}{l}
{\footnotesize 
$^{(a)}$ The contributions from $~{\cal L}_{1,2,13}~$ are always associated
with a factor of $~\sin^2\theta_W~$, unless specified otherwise.}\\
{\footnotesize 
~~~$~{\cal L}_{4,5,6,7,10}~$ do not contribute to the 
processes considered in this table. 
Also, for contributions to the $B$-term}\\
{\footnotesize 
~~~~in a given $V_L$-amplitude, 
we list them separately with the $B$-term specified.}\\
{\footnotesize  $^{(b)}$ MI~$=$~model-independent, MD~$=$~model-dependent.
~~}\\
{\footnotesize  $^{(c)}$ Here, $~B_0^{(0)}~$ is dominated by 
$~T_0[q\bar{q};2v]~$ since $~T_0[q\bar{q};\pi,v]~$ contains a 
suppressing factor $\sin^2\theta_W$ as can be}\\
{\footnotesize 
~~~~deduced from $~T_0[q\bar{q};\pi V_T]~$ 
times the factor $~v^\mu =O\left(\frac{M_W}{E}\right)~$.}\\
\end{tabular}
\end{center}
\end{table}

\end{document}